\documentstyle[pra,aps, psfig]{revtex}

\begin{document}
\title{ Using atomic interference to probe atom-surface interaction}
\author{Roberta Marani, Laurent Cognet, Veronique Savalli, Nathalie
Westbrook, Christoph I. Westbrook, Alain Aspect}
\address{Laboratoire Charles Fabry de l'Institut d'Optique, Unit\'e
Mixte du CNRS n$^\circ$ 8501, BP
147, 91403 Orsay CEDEX, France}

\date{\today}
\maketitle

\begin{abstract}
We show that atomic interference in the reflection 
from two suitably polarized evanescent waves is sensitive to retardation effects in
the atom--surface interaction for specific experimental parameters.
We study the limit of short and long atomic de Broglie wavelength. The
former case is analyzed in the semiclassical approximation
(Landau-Zener model). The latter represents a quantum regime and is
analyzed by solving numerically the associated coupled
Schr\"odinger equations. We consider a specific experimental scheme
and show the results for rubidium (short wavelength) and the much lighter
meta-stable helium atom (long wavelength). The merits of each case are
then discussed. 
\end{abstract}
\pacs{ }

\section{Introduction}
\label{sec:intro}

The interaction between a ground--state atom 
and a dielectric or conducting
wall has been investigated theoretically (\cite{lennard:32},
\cite{casi-pold:48},\cite{dzya-etal:61},\cite{spru-tiko:93},
\cite{tiko-spru:93},
\cite{spruch:96}, \cite{fich-etal:95}, \cite{wyli-sipe:84},
\cite{wyli-sipe:85}, \cite{mavroyann:63}) and experimentally 
(\cite{land-etal:96},
\cite{kase-etal:91}, \cite{sand-etal:92}, \cite{suke-etal:93}). 
Theoretical studies have been performed on different levels,
from a simple model of a dipole--dipole interaction of the atom and its
mirror image, to the full QED relativistic quantum
treatment. Interesting in particular are the long--range Casimir
interactions \cite{spruch:96} that were recently observed in cavity QED
experiments \cite{sand-etal:92}, \cite{suke-etal:93}. 
When the atom--wall distance  $z$ is not small compared to the
wavelength of the dominant atomic transitions, the $z^{-3}$ law
associated with the instantaneous electrostatic interaction is no
longer valid. The full QED treatment leads to the famous long distance
$z^{-4}$ law. 

Recent experimental developments enable precise manipulation of cold
atoms by lasers, see e.g. Ref.  \cite{nobel:97}.
Small and well defined velocities of the
atoms can be achieved using advanced cooling and launching techniques
and a detuned laser field can be used to create controlled and
adjustable potentials for the atoms. Under these conditions, atoms can
be used to explore the atom--surface potential, for example using
evanescent--wave mirrors.
Classical reflection
from such an atomic mirror was used to measure the van der Waals force
between a dielectric surface and an atom in its ground state
\cite{land-etal:96}. This experiment though, could not fully
discriminate between the electrostatic and the QED expressions.
Segev {\it et al.} \cite{sege-etal:97} considered a similar experiment 
in the quantum
regime (atoms undergoing above--barrier, classically forbidden
reflection). Unlike classical reflection, which can only be
used to identify thresholds and to measure the height of the potential
barrier, quantum
probabilities are determined by the complete potential curve, and are
sensitive to the short- and long--range behavior of the
potential. 
It was found that, for velocities of the order of the recoil speed, 
the quantum reflection probabilities are indeed
sensitive to the long--range (Casimir) interaction.

In this work we study how the form of the atom--surface
interaction can be observed using atomic interference of the type
reported in \cite{cogn-etal:98}. We consider
atoms with multiple ground state sub-levels, which feel different
potentials in the evanescent radiation field. These potentials can be
probed by using stimulated Raman transitions within the evanescent
wave. These transitions exhibit interference effects whose phases
depend on the atomic trajectories and on the entire potential, as in
quantum reflection. An important aspect of the effects we discuss here
is that they occur for higher incident velocities than those considered
in \cite{sege-etal:97} and may therefore be experimentally easily
accessible.
Furthermore, even for small velocities (near the recoil velocity) the effects
are more dramatic than those shown by quantum reflection.

As discussed in Refs.\cite{henk-etal:97} and \cite{cogn-etal:98}, 
the experiment is analogous to an atom interferometer whose size is
of the order of a fraction of the evanescent wave decay length, $\kappa^{-1}$.
We  will focus on three experimental cases:
in the first, the atomic de Broglie wavelength $\lambda_{\rm dB} \ll \kappa^{-1}$.
This leads to interference fringes analogous to those in any other 
interferometer.
It corresponds to the experiment described in Ref. \cite{cogn-etal:98}
in which rubidium (Rb) atoms are dropped from a height of about 2 cm.
Second, we will repeat the situation of the first case, 
but with Rb replaced by meta-stable helium (He*). 
In this case we have $\lambda_{\rm dB}/2\pi \simeq$ a few
$\kappa^{-1}$, which allows for only a few interference fringes.
This allows us to illustrate how the signal changes when only
the atomic species is changed. 
Finally, we will examine the case where 
$\lambda_{\rm dB}/2\pi \simeq \kappa^{-1}$, 
taking He* atoms at the recoil velocity as our example. 
No fringes are present, but a strong dependence on the nature of
the atom-wall potential is demonstrated.
To simplify the discussion, we work with a two-level model,
between the initial atomic state and the adjacent one in the ground 
state manifold, thus neglecting
the populations of the other ground state sub-levels as well as the 
excited levels.
For the atom-wall interaction we will use the published values for each atom.

In the short $\lambda_{\rm dB}$ limit, the motion of the atom
can be treated semi-classically. We thus calculate the
transition probability between the two atomic levels according to the
Landau-Zener model for adiabatic transitions.
In the other cases,
the study of the atomic motion requires a fully quantum mechanical 
treatment. We
study these cases by solving the associated coupled Schr\"odinger
equations numerically. 
 
In Sec.\ref{sec:2} we describe the physical system and the theory for
atomic interference. In Sec.\ref{sec:res} the results are presented.
Discussion and conclusions are given in Sec.\ref{sec:end}.

\section{Model for interference of multiple ground state sub-levels}
\label{sec:2}

\begin{figure}
\centerline{\psfig{figure=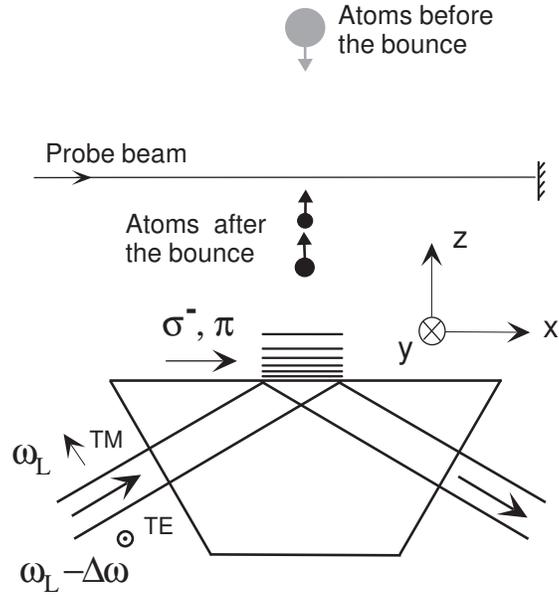,height=8cm}}
\caption{Typical experimental setup. The atoms are released from a MOT
above the prism. Two slightly detuned laser beams of polarizations TM
(and frequency $\omega /(2\pi)$) and TE 
(and frequency $(\omega  - \Delta\omega)/(2\pi)$)
form evanescent waves on the surface of the prism of polarizations
$\sigma^-$ and $\pi$ respectively ($y$ is the quantization axis).
The reflected atoms
are detected by a detection beam.}
\label{fig:setup}
\end{figure}

We will focus on an
experimental setup along the lines of Fig. \ref{fig:setup}.
Here, a strong laser beam with frequency $\omega $ and transverse
magnetic
(TM) polarization
(i.e. magnetic field perpendicular to the plane of incidence) creates
an evanescent wave, which is nearly $\sigma^-$ circularly polarized with
respect to the y-axis.
A second weak laser beam with frequency $\omega  - \Delta\omega$ and
transverse electric (TE)
polarization (i.e. electric field perpendicular to the plane of
incidence) creates another evanescent wave with $\pi$ polarization
along the y-axis.

Atoms normally incident on the evanescent wave move in an
effective optical potential $\hat V_{\rm light}$ 
(which in general depends on the internal
state of the atom) and an attractive atom--wall interaction 
$\hat V_{\rm wall}$. 
The total potential is given by 
\begin{equation}
\label{eq:U}
\hat U(z)=\hat V_{\rm light}(z) + \hat V_{\rm wall}(z).
\end{equation}

\subsection{Optical potential}

Let us consider an atom in the evanescent field and assume for
simplicity that the   $\sigma^-$ polarized wave is strong compared to
the $\pi$ wave. The $\sigma^-$
component lifts the magnetic (Zeeman) degeneracy of the atomic
ground state (the
quantization axis is in the $\hat y$ direction), so that each magnetic
sub-level feels a different optical potential.
To first order, the $\pi$
polarized wave produces a coupling between the atomic sub-levels via 
a Raman transition: for example,
starting from the sub-level $m_i$, the atom
absorbs a $\sigma^-$ polarized photon and emits a stimulated photon with $\pi$
polarization. The atom thus ends
up in the $m_i -1$ sub-state with its total energy increased by 
$\hbar\Delta\omega$. 
This transition is resonant when 
$\hbar\Delta\omega$ is equal to the energy difference between the
magnetic sub-levels. 
If the two evanescent waves are counter-propagating, in the reference
frame moving with the optical grating, the situation 
corresponds to grazing incidence diffraction \cite{cogn-etal:98}.  
For a review of the theoretical understanding of atomic
diffraction and interference from evanescent waves, see
\cite{henk-etal:}. 

In the limit of low saturation and a detuning $\delta$ large compared to
both the frequency difference $\Delta\omega$ and the natural linewidth,
the excited state
manifold may be eliminated adiabatically, and, for a ground state
of total angular momentum (including nuclear spin) $J_g$, the atomic
wave function is described by the
$2J_g+1$ Zeeman components $|m>, m=-J_g,...,+J_g$. An atom at a distance $z$
from the surface of the mirror is subject to
an optical potential $\hat V_{\rm light}(z)$ whose matrix elements are
of the form
[we suppose that the frequency difference $\Delta\omega$ and Zeeman shift are
negligible compared to the detuning $\delta$ and the hyperfine
structure of the excited level]
\begin{equation}
\label{eq:Vdip}
<m|\hat V_{\rm light}(z)|m'>=\frac{d^2}{\hbar \delta}\sum_{q,q',m_e, J_e}
E_q^*(z)E_{q'}(z) (J_g m; 1 q | J_e m_e )(J_e m_e | J_g m'; 1 q'),
\end{equation}
where a product of Clebsch--Gordon--coefficients appears on the rhs,
the electric field polarization is expanded in the usual spherical basis with
coefficients $E_q$, ($q=-1,0,+1$) and $d$ is the reduced dipole moment. 
The optical potential couples different
Zeeman sub-levels if the field is not in a pure polarization state
with respect to this basis, as in the setup considered in this work. 
The optical potential due to the two evanescent waves can then be written as
\begin{equation}
\label{eq:Vdip2}
<m|\hat V_{\rm light}(z)|m'> = \frac{d^2}{\hbar\delta} C_{mm'} 
\exp(-2 \kappa z),
\end{equation}
where the $C_{mm'}$ coefficients are given by the Clebsch-Gordon
coefficients times the field amplitudes at the surface $z=0$ 
(see Eq.~\ref{eq:Vdip}).
The inverse decay length $\kappa$ is 
$\kappa = k [ (n \sin \theta)^2 - 1 )]^{1/2}$,
where $k=\omega /c$ is the free field wave vector, assumed to be the same for
each laser, $n$ is the refraction index of the prism and $\theta$ is
the angle of incidence of the lasers with the surface.
In our calculations we will always use $n=1.87$ and $\theta = 53^\circ$.
The wave-vector $k$ is different for each atom.

\subsection{Atom--wall potential}

The simplest model for the interaction of a ground state atom and a
wall of dielectric constant $\epsilon$ considers the interaction
between a dipole {\bf d} and its mirror image and yields the
Lennard--Jones potential,
\begin{equation}
V_{\rm wall}^{\rm LJ}(z)=- \frac{\epsilon - 1}{\epsilon + 1}
\left(\frac{<d_{\parallel}^2> + 2<d_{\perp}^2>}{64\pi\epsilon_0}\right) 
\frac{1}{z^3} 
\equiv - \frac{\epsilon - 1}{\epsilon + 1} \frac{C^{(3)}}{z^3},
\label{eq:VLJ}
\end{equation}
where $<d_{\parallel}^2>$ and $<d_{\perp}^2>$ are the expectation
values of the squared dipole parallel and perpendicular to the surface
\cite{lennard:32},\cite{cote-etal:98}. This expression for the potential is
approximately valid for $\epsilon$ independent of frequency 
and  $kz$ much smaller than unity.  

If we take into account retardation effects, the Casimir-Polder
potential is obtained, where the finite propagation time between the
dipole and its image results in a different power-law
behavior for large $z$\cite{casi-pold:48},

\begin{equation}
\lim_{z \to \infty} V_{\rm wall}^{CP}(z) \propto z^{-4}.
\end{equation}

In the complete QED theory the interaction potential between an atom of
polarizability $\alpha(\omega)$ at a distance $z$ from a dielectric
wall can be written as
\cite{tiko-spru:93}:
\begin{equation}
V_{\rm wall}^{\rm QED}(z)=- \frac{\hbar}{8\pi^2c^3} \int_0^{\infty} d\omega 
\omega^3 \alpha(\omega) \times \left(\int_0^1 dp+\int_0^{i\infty} dp
\right) H(p,\epsilon)\exp(-2ip\omega z/c)
\end{equation}
with
\begin{equation}
H(p) = \frac{\sqrt{\epsilon - 1 + p^2} - p}{\sqrt{\epsilon - 1 + p^2}
+ p} + (1-2 p^2) \frac{\sqrt{\epsilon - 1 + p^2} - \epsilon
p}{\sqrt{\epsilon - 1 + p^2}+\epsilon p}.
\nonumber
\end{equation}

The
numerical values of the constant $C^{(3)}$ for Rb and He* atoms used in
this paper are given in Table \ref{tab:1}, as well as the numerical values
of some other
important parameters. Since the
atom is in a ground state with $l=0$, the value of $C^{(3)}$ is the same
for any magnetic or hyperfine sub-level.
The detailed interaction between a ground state
Rb or meta-stable He atom and a dielectric surface were recently
calculated.
For the van der Waals potential we have used data from \cite{landragin:97}
for Rb and from  \cite{yan-babb:98} for He*. In the former work an 
interpolation formula for the van der Waals potential 
is given as $V^{QED}_{\rm wall}(z)=V^{LJ}_{\rm wall}(z)0.987[(1+1.098 z)^{-1} -
0.00493 z (1 + 0.00987 z^3 - 0.00064 z^4)^{-1}]$ where $z$ is
expressed in units of the laser wavelength $\lambda/(2\pi)$. This formula
approximates the numerical calculation with a 0.6\% accuracy between 0
and $10\lambda/(2\pi)$.

\subsection{Population transfer}

\begin{figure}
\centerline{\psfig{figure=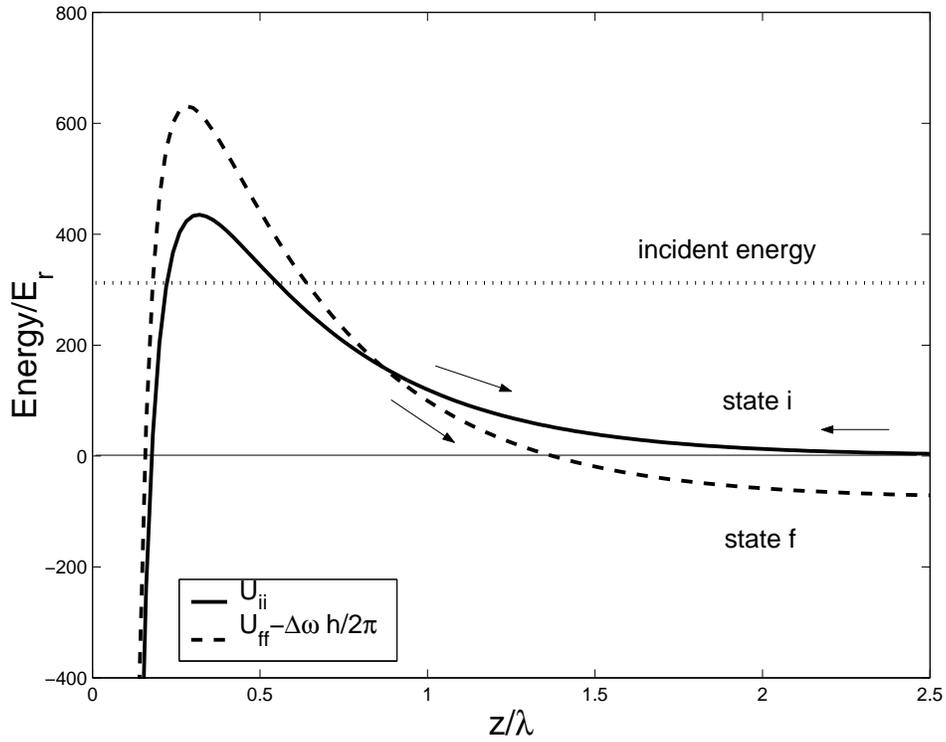,height=10cm}}
\caption{Typical potential curves experienced by the atoms during
reflection. The potential for state $f$ is shifted by the Raman energy
$-\hbar\Delta\omega$. The atoms approach the potential barrier in 
state $i$, pass through the curve crossing twice and can end up
on either state $i$ or $f$. For each final situation, two paths are
possible (between the crossing and the turning points) and can interfere
producing fringes as a function of the location of the crossing.}
\label{fig:pot}
\end{figure}

In this work we will consider only population
transfer from the incident sub-level $i$ to the final one $f$
($m_f=m_i-1$), that is only two
levels. This is a good approximation if the coupling is weak enough
for the population of the other levels to be negligible.
An example of the total interaction potential from Eq.~\ref{eq:U} 
is shown in Fig.\ref{fig:pot}. The $U_{ff}$ potential curve has been shifted
vertically by $-\hbar\Delta\omega$, corresponding to the kinetic energy change.
Then the two (adiabatic) potential curves cross at the point of
resonance. The coupling turns the exact crossing into an avoided crossing.
An incoming wave function in the
$m_i$ channel is split in two parts that are subsequently
reflected from their respective repulsive potentials and recombined
after the second passage at the crossing. 
Thus, the evanescent wave realizes a ``Michelson
interferometer'' with a single beam splitter and two mirrors. 

When the atomic $\lambda_{\rm dB}$ is short,
the avoided crossing can be treated by means of the semi-classical 
Landau--Zener model
for non--adiabatic transitions \cite{zener:32}.
Assuming that the atom moves through the crossing with a constant
velocity $v_c$ (fixed by energy conservation), the Landau--Zener formula
allows one to compute the probability amplitude for the two
atomic levels after the crossing. 
The transition probability $w_{if}$ from the initial sub-level $i$ to the final
sub-level $f$ is given by
\begin{equation}
\label{eq:LZ}
w_{i,f}=4 T_{LZ} R_{LZ} \cos^2(\delta\phi),
\end{equation}
where $R_{LZ}=1-T_{LZ}$ and the transmission coefficient
is
$
T_{LZ}=\exp(-\pi\Lambda)
$,
with
$
\Lambda={2|<i|\hat U|f>|^2}/(\hbar^2\kappa\Delta\omega v_c).
$

The phase difference $\delta\phi$ is given by the difference in 
the phase shifts between
the crossing and the turning points in the semi-classical
approximation, plus a correction term \cite{kaza-etal:85}: 

\begin{equation}
\delta\phi= \frac{1}{\hbar}\left[\int_{z_{f,r }}^{z_c}dz p_f(z)-
\int_{z_{i,r }}^{z_c}dz p_i(z)\right]+
\frac{\pi}{4}+\frac{\Lambda}{2}\log\left(\frac{\Lambda}{2e}\right) + 
\arg\left(\Gamma\left(1-i\frac{\Lambda}{2}\right)\right),
\end{equation}
where $z_c$ is the position of the crossing point, $z_{n,r}$ the
classical return point for an atom in the n-th level, $p_n(z)$ its momentum
and $\Gamma$ the Gamma function.

Changing the frequency difference $\Delta\omega/(2\pi)$ causes a
change in the length
of one of the interferometer arms, thus a change in the phase difference
$\delta\phi$ between the two paths.
As a consequence we expect the transition probability to show
oscillations in  $\Delta\omega$ (St\"uckelberg oscillations).  
We will see in the next section that $\delta\phi$ is very sensitive to
the exact shape of the potential.
The amplitude of the oscillations also depends on
$\Delta\omega$ both explicitly and implicitly through $v_c$
(the crossing point moves with changing 
$\Delta\omega$).

The Landau--Zener model is a good approximation only when the atom
speed is approximately constant during the crossing. In particular it
is not valid when the classical return point and the crossing are
close to each other or when the de Broglie wavelength of the atom is
of the order of the width of the interaction region. 
In order to explore this long-wavelength regime, we have to forgo the
semi-classical Landau-Zener model and
solve numerically the coupled Schr\"odinger equations for the system.
Since atoms that cross the potential barrier stick to the
dielectric surface, the appropriate boundary conditions at the surface
are those for a running wave propagating downward ($z\to -\infty$), while
for $z\to\infty$ the solution is a superposition of downward
(incident) and upward (reflected) waves.
We have integrated the system of Schr\"odinger equations using the
renormalized Numerov method \cite{johnson:77}. To avoid the
singularity at the surface we have modified the potential to be a
large negative constant near the interface and verified that the
transition probability does not change by varying the value of the constant.

\section{Results}
\label{sec:res}

We will present our calculations in two parts.
First we discuss the short wavelength regime, and point
out various experimental strategies to observe
retardation effects. 
Then we discuss two cases in which the de Broglie wavelength is not
small compared to the evanescent wave decay length 
(the "long wavelength regime"), 
in which a numerical integration of the Schr\"odinger equation
is necessary.

\subsection{Short wavelength regime}

\begin{figure}
\centerline{\psfig{figure=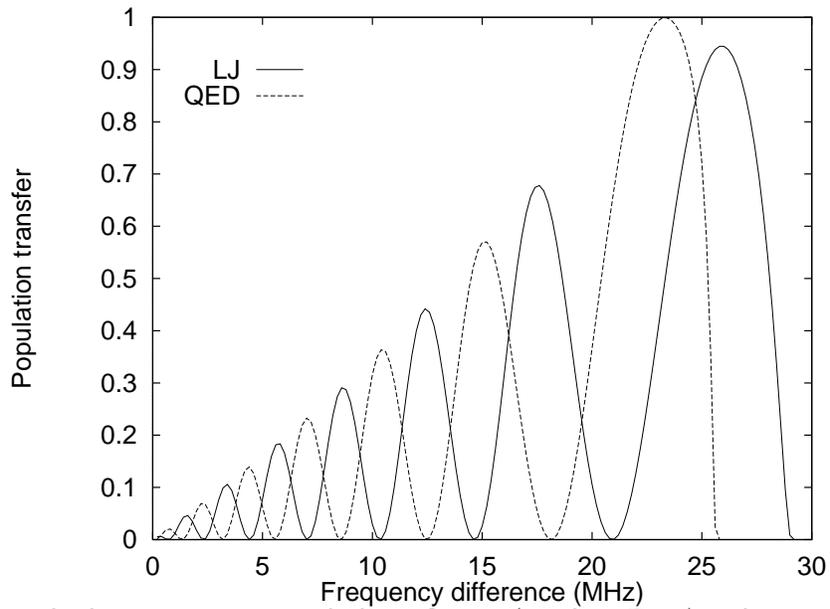,height=8cm}}
\caption{Population transfer from state $i$ to state $f$ after
reflection ($w_{if}$ from Eq.~\protect\ref{eq:LZ}) vs frequency difference 
$\Delta\omega/(2\pi)$ for rubidium atoms, released 2.3 cm
above the mirror, (i.e. with incident momentum $115 \hbar k $). 
The solid line is
for the QED model and the dashed line for the Lennard--Jones (LJ)
model.
The coupling coefficients defined in Eq.~\protect\ref{eq:Vdip2} are 
$|C_{if}|=859 E_r^{\rm Rb}$ and $|C_{ii}|=3.8\times 10^4 E_r^{\rm Rb}$}.
\label{fig:Rb1}
\end{figure}

Fig.~\ref{fig:Rb1} shows a calculation of the
population transfer $w_{if}$ as a function of the frequency difference $\Delta
\omega$ for the Lennard--Jones (LJ) and QED model of the van der Waals
interaction. 
The value of the
light potential is the same for the two curves (i.e. the laser
intensity is the same), and the incident momentum is $115 \hbar k$,
which corresponds to Rb atoms dropped from a height of 2.3 cm. 
We use the Landau-Zener approach to calculate the transfer
probability (see Eq. \ref{eq:LZ}). As in \cite{henk-etal:97} and
\cite{cogn-etal:98}, we observe St\"uckelberg
oscillations in the transfer probability. These oscillations can be
understood as the variations in the accumulated phase difference between the
two different paths taken by the atoms after the level crossing shown in
Fig.2. The de Broglie wavelength $\lambda_{\rm dB}$ 
is such that several fringes appear as the
position of the level crossing is moved through its possible range.
The
last oscillation at the higher frequencies is where the crossing point
and the classical return point are very close to each other and the
Landau-Zener model breaks down. We have set the probability to zero 
beyond this limit. In reality, the transition probability falls roughly
exponentially with frequency, as one finds solving the Schr\"odinger
equations numerically (see next subsection). 

\begin{figure}
\centerline{\psfig{figure=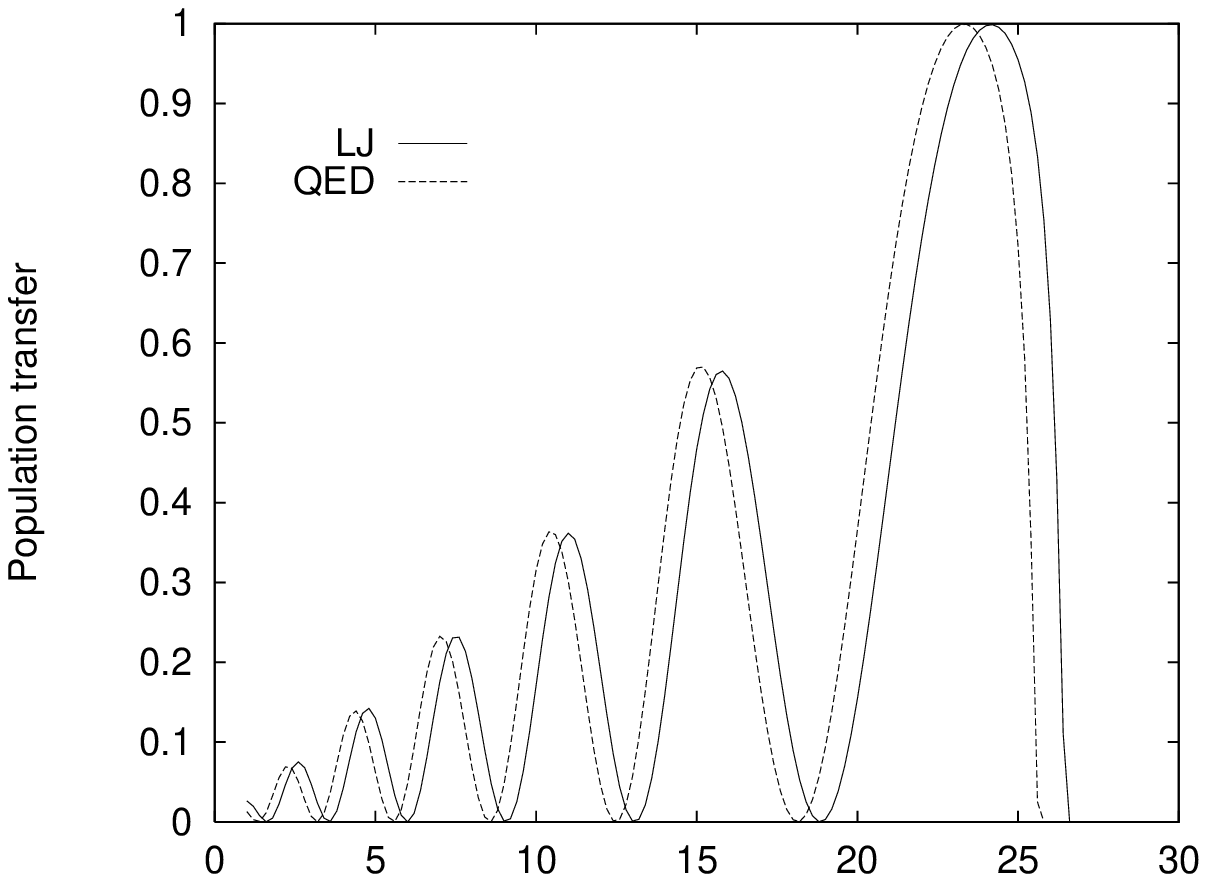,height=8cm}}
\centerline{\psfig{figure=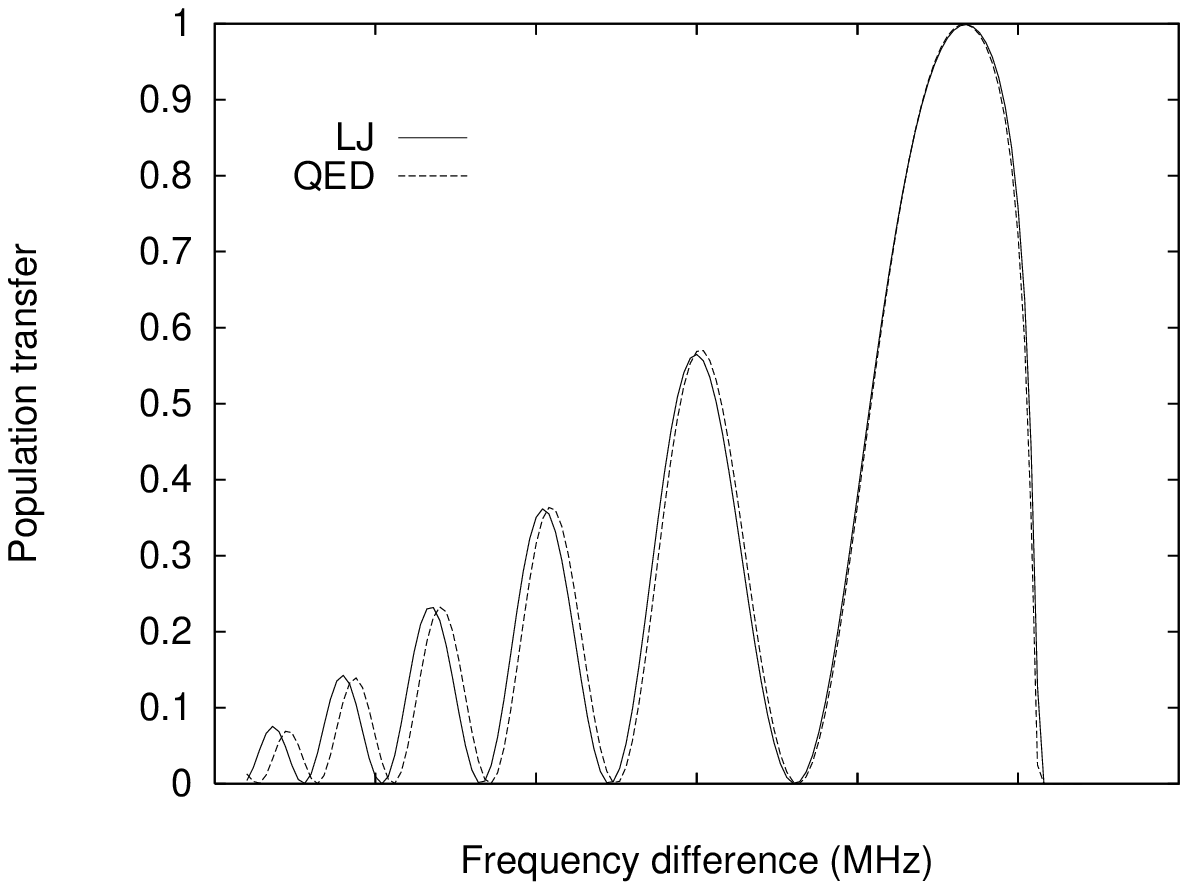,height=8cm}}
\caption{Upper figure: population transfer $w_{if}$ (see
Eq.~\protect\ref{eq:LZ}) vs frequency difference $\Delta\omega/(2\pi)$ for 
rubidium atoms released 2.3 cm
above the mirror (i.e. with incident momentum $115 \hbar k_{\rm Rb}$). 
The solid line is for the QED model and the dashed line for the
Leenard--Jones (LJ) model.
The height of the potential barrier is 
$1.48\times 10^4 E_r^{\rm Rb}$, i.e. the
intensity of the strong laser beam has been adjusted to give the same barrier
height for both models.
The light--shift coefficients for the QED model are the same as for
Fig.~\protect\ref{fig:Rb1}, while for the LJ model
$|C_{ii}|=4.11\times 10^4 E_r^{\rm Rb}$ and 
$|C_{if}|=890 E_r^{\rm Rb}$.
Lower figure: same parameters as in the previous figure with the LJ
curve artificially shifted 0.79 MHz to the left in order to show the
changes in fringe spacing.}
\label{fig:Rb2}
\end{figure}

In general, we find that the dephasing between the two
curves (with and without retardation effects) is
greatest when the atoms are incident at an energy close to the top of
the potential barrier.
Note that the height of
the barrier is greater when retardation effects are included, since
these reduce the strength of the atom-surface interaction.
The effect of retardation is roughly to shift these fringes by half a fringe
to the left. 
The major cause of this shift is the increase in the height
of the total potential which is greater for the $i$ level than the $f$ level.
A 10\% increase in the value of the
light-shift potential would exhibit nearly the same shift. Therefore, since
it is not possible to turn retardation on and off, it would be necessary to
measure absolutely the light-shift potential to better than 10\% in
order to distinguish a retardation effect. Experimentally this is 
rather difficult.
Instead of attempting to measure the absolute light shift, one could rather
measure the absolute height of the potential by observing the
threshold of reflection as in \cite{land-etal:96}, and using the known kinetic
energy of the atoms to get an absolute calibration of the height. 
Let us assume then that the barrier height, instead of the light intensity, is
known.
In Fig.\ref{fig:Rb2}a we show the result
with the same parameters as in Fig.\ref{fig:Rb1} except for the light intensity
in the LJ model, which has been changed so as to have the same barrier
height as the QED model. 
In this case
the shift is much smaller, about 1/5 of a fringe.
We have verified that even taking into account an experimental
uncertainty of a few \% in the height
of the corresponding potentials, the two models are still clearly
resolved. 

This approach seems feasible, but a third method of observing the effects
of retardation is possible if one uses more of the information available in
the oscillation curve. Fig. 4b shows the same curves as Fig. 4a but with the
QED curve numerically shifted so as to coincide with the LJ curve at 
its maximum. One
sees that the period of these oscillations is not the same. 
It decreases with decreasing detuning, faster for the
full QED potential, so that there is a difference in the spacing of the 
minima in the population transfer. Thus with
fringes with sufficiently high signal to noise, one can distinguish
retardation while leaving the absolute barrier height as a free parameter
in a fit to the data. It seems to us that a viable experimental method is
to use a combination of the second two approaches. Careful measurements of
the barrier height can be used to cross check a fit to the St\"uckelberg
oscillations with the barrier height as a free parameter.

The incident energy, momentum and barrier height used in Figs. 3 and 4a were
arbitrarily chosen to correspond to the experiment in
\cite{cogn-etal:98}, but it
would be interesting to know how Figs. 3 and 4a would change, especially for
different incident momenta (de Broglie wavelengths). We studied this question
by repeating the calculation of Fig. 4a, for different incident momenta,
while always keeping the barrier height 10\% above the incident kinetic
energy. We find that, roughly speaking, the number of oscillations
increases as the incident momentum. This is because the number of fringes
in the interferogram increases with decreasing wavelength. The accumulated
phase difference between the oscillations in the LJ and QED models,
over the corresponding frequency range, also
increases approximately linearly with incident momentum. Thus if we
consider the fringe shift divided by the fringe period as a figure of
merit, the sensitivity of the experiment to retardation effects increases
with increasing incident momentum.

\subsection{Long wavelength regime}

\begin{figure}
\centerline{\psfig{figure=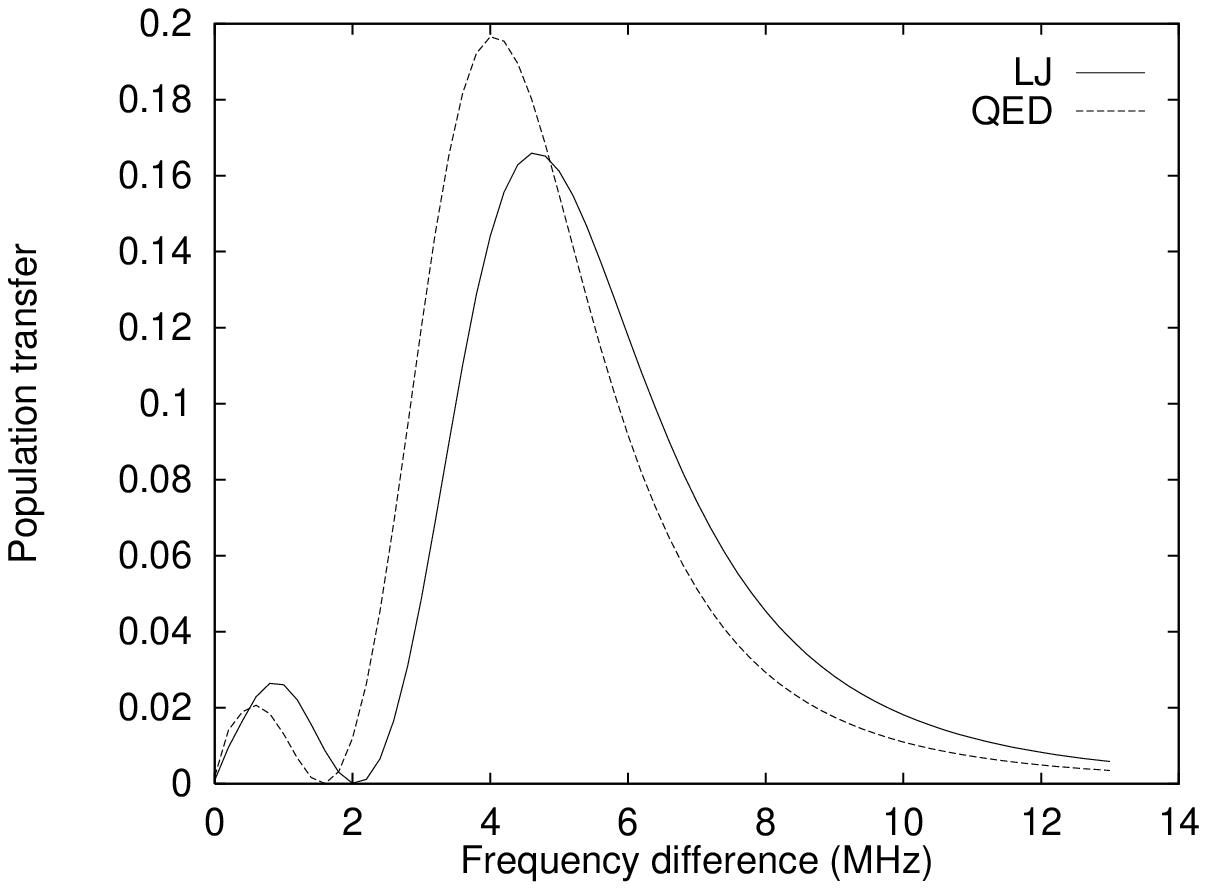,height=8cm}}
\caption{Population transfer $w_{if}$ (see Eq.~\protect\ref{eq:LZ}) vs 
frequency
difference $\Delta\omega/(2\pi)$ for meta-stable helium atoms released 2.3 cm
above the mirror (i.e. with incident momentum $7.4 \hbar k_{\rm
He}$).
The solid line is
for the QED model and the dashed line for the Lennard--Jones (LJ) model.
The height of the potential barrier is 
$61 E_r^{\rm He}$, i.e. the
intensity of the strong laser beam has been adjusted to give the same barrier
height for both models ( 10\% above the incident kinetic energy).
The light--shift coefficients for the QED model are 
$|C_{ii}|=252 E_r^{\rm He}$
$|C_{if}|=20.8 E_r^{\rm He}$
while for the LJ model
$|C_{ii}|=303 E_r^{\rm He}$ and
$|C_{if}|=22.7 E_r^{\rm He}$.}
\label{fig:He1}
\end{figure}

We now examine the large de Broglie wavelength limit,
where the semi-classical
model breaks down since the atomic wavelength is not small compared to the
interferometer size.
We first consider the case of He* atoms dropped from a height
of a few cm as for Rb. We show the results in Fig.~\ref{fig:He1}. 
For an initial distance of 2.3 cm from the mirror, the incident momentum of He*
is $7.4 \hbar k$, much lower than for Rb. 
Again we will chose the intensities of the strong $\sigma^-$ wave so 
as to have the
same barrier height for the two potentials, about 10\% above the
incident kinetic energy.
In this regime, to observe interference, the detuning between the
evanescent waves must be of only a few MHz, in fact, beyond about 10MHz
the crossing point is closer to the surface than the return point.
One only sees one or two St\"uckelberg oscillations since the momentum
involved is small and the atoms do not accumulate enough phase difference
to show more oscillations. 
The two potentials give similar results (Fig.\ref{fig:He1}), the main
difference being in the shape and height of the big peak.

\begin{figure}
\centerline{\psfig{figure=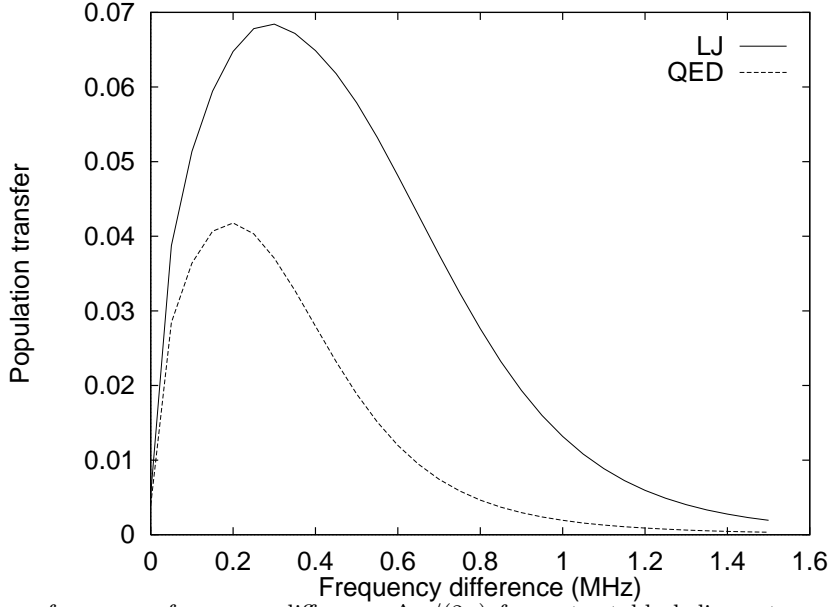,height=8cm}}
\caption{Population transfer $w_{if}$ vs frequency
difference $\Delta\omega/(2\pi)$ for meta-stable helium atom with
incident momentum $\hbar k_{\rm He}$.
The solid line is
for the QED model and the dashed line for the Lennard--Jones (LJ) model.
The height of the potential barrier is 
$1.28 E_r^{\rm He}$, i.e. the
intensity of the strong laser beam has been adjusted to give the same barrier
height for both models.
The light--shift coefficients for the QED model are 
$|C_{ii}|=0.8 E_r^{\rm He}$
$|C_{if}|=0.16 E_r^{\rm He}$
while for the LJ model
$|C_{ii}|=1.47 E_r^{\rm He}$ and
$|C_{if}|=0.29 E_r^{\rm He}$.} 
\label{fig:He2}
\end{figure}

We have also looked at even lower incident momenta.
Near the recoil limit, one can expect interference only for detunings of
less than one MHz and the accumulated phase difference is too small in
this range to show any St\"uckelberg oscillations.
Both for He* and Rb one
obtains a big peak in the transition probability and no St\"uckelberg
oscillations, as expected. The difference between the van der Waals
potential and the full QED potential still shows up in the different
shape and height of the peaks (Fig.\ref{fig:He2}). 
Here the width of the curve is delimited by the frequency for which
the classical return point and the crossing point coincide.

Finally we note that the qualitative features of the short and long
$\lambda_{\rm dB}$ limit are the same for Rb and He*, e.g. He* with an
incident momentum of about $100\hbar k$ gives the same type of
oscillation pattern as Rb.

\section{Conclusions}
\label{sec:end}

In summary, we have proposed an experiment to probe van der Waals like
surface interactions by exploiting interference mechanisms for
well defined Zeeman sub-levels of atoms moving in two evanescent
waves.
Retardation can be
resolved using atoms incident at speeds which are easily
obtained in free fall over a few centimeters. The controlling parameter is
here the detuning between the two evanescent waves.
One then measures the fraction of the atoms which have undergone a
change of magnetic sub-level as a function of this detuning.

For the situation in which the atomic de Broglie wavelength is 
sufficiently small
the experiment resembles typical interferometry experiments. 
The theoretical description is semi-classical, employing
well defined atomic trajectories, while experimentally, one seeks
a particular (non sinusoidal) fringe pattern as a signature of retardation. 
This should be possible with an improved version of the experiment
of Ref. \cite{cogn-etal:98}. 

Another approach is to investigate the interaction of atoms 
whose de Broglie wavelength is not small compared to
the length scale of the interferometer. 
In this regime most of the information is to be found in the shape of 
the population transfer curve, since there are very few or zero 
interference fringes.
Note though, that we have assumed throughout that the incident atoms
are mono-energetic. This means that the velocity spread of 
the incident atoms must be small compared to the 
atomic recoil.
Nevertheless this may be worth the effort, 
because the predicted effect of retardation is quite
dramatic.

For a quantitative comparison however, the presence of all the
sub-levels (which give rise to the multiple crossing of the dressed
potentials at the same distance $z_c$ from the surface)
and, possibly, losses from spontaneous emission have to be
taken into account, but the results are not qualitatively different.

\section{Acknowledgments}
R. M. acknowledges support from the Training and Mobility of
Researchers 
(European Union, Marie--Curie Fellowship contract n. ERBFMBIC983271)
and would like to thank Paul Julienne and
Olivier Dulieu for their help with the numerical code, and J. Babb for
providing the data for the atom-wall interaction for meta-stable helium.
This work was also supported by the R\'egion Ile de France.

%\bibliographystyle{prsty}
%\bibliography{mirror}

\begin{thebibliography}{10}

\bibitem{lennard:32}
L.~E. Lennard-Jones, Trans. Faraday Soc. {\bf 28},  333  (1932).

\bibitem{casi-pold:48}
H.~B.~G. Casimir and D. Polder, Phys. Rev. {\bf 73},  360  (1948).

\bibitem{dzya-etal:61}
I.~E. Dzyaloshinskii, E.~M. Lifshitz, and L.~P. Pitaevskii, Adv. Phys. {\bf
  10},  165  (1961).

\bibitem{spru-tiko:93}
L. Spruch and Y. Tikochinsky, Phys. Rev. A {\bf 48},  4213  (1993).

\bibitem{tiko-spru:93}
Y. Tikochinsky and L. Spruch, Phys. Rev. A {\bf 48},  4223  (1993).

\bibitem{spruch:96}
L. Spruch, Science {\bf 272},  1452  (1996).

\bibitem{fich-etal:95}
M. Fichet, F. Schuller, D. Bloch, and M. Ducloy, Phys. Rev. A {\bf 51},  1553
  (1995), and references therein.

\bibitem{wyli-sipe:84}
J.~M. Wylie and J.~E. Sipe, Phys. Rev. A {\bf 30},  1185  (1984).

\bibitem{wyli-sipe:85}
J.~M. Wylie and J.~E. Sipe, Phys. Rev. A {\bf 32},  2030  (1985).

\bibitem{mavroyann:63}
C. Mavroyannis, Mol. Phys. {\bf 6},  593  (1963).

\bibitem{land-etal:96}
A. Landragin {\it et~al.}, Phys. Rev. Lett. {\bf 77},  1464  (1996).

\bibitem{kase-etal:91}
M. Kasevich {\it et~al.},  in {\em Atomic Physics 12}, Vol.~233 of {\em AIP
  Conf. Proc. No. 233}, edited by J.~C. Zorn and R.~R. Lewis (AIP, New York,
  1991), p.\ 47.

\bibitem{sand-etal:92}
C.~I. Sandoghdar, V.~Sukenik, E.~A. Hinds, and S. Haroche, Phys. Rev. Lett.
  {\bf 68},  3432  (1992).

\bibitem{suke-etal:93}
C.~I. Sukenik {\it et~al.}, Phys. Rev. Lett. {\bf 70},  560  (1993).

\bibitem{nobel:97}
S. Chu, C.~N. Cohen-Tannoudji, and W.~D. Phillips, Rev. Mod. Phys. {\bf 70},
  685  (1998).

\bibitem{sege-etal:97}
B. Segev, R. C\^ot\'e, and M.~G. Raizen, Phys. Rev. A {\bf 56},  R3350  (1997).

\bibitem{cogn-etal:98}
L. Cognet {\it et~al.}, Phys. Rev. Lett. {\bf 81},  5044  (1998).

\bibitem{henk-etal:97}
C. Henkel, K. Molmer, R. Kaiser, and C.~I. Westbrook, Phys. Rev. A {\bf 56},
  R9  (1997).

\bibitem{henk-etal:}
C. Henkel {\it et~al.}, Appl. Phys. B {\bf 69}, 277 (1999).

\bibitem{cote-etal:98}
R. C\^ot\'e, B. Segev, and M.~G. Raizen, Phys. Rev. A {\bf 58},  3999  (1998).

\bibitem{landragin:97}
A. Landragin, Ph.D. thesis, Universit\'e de Paris--Sud, 1997.

\bibitem{yan-babb:98}
Z.-C. Yan and J.~F. Babb, Phys. Rev. A {\bf 58},  1247  (1998).

\bibitem{zener:32}
C. Zener, Proc. Roy. Soc., Ser. A {\bf 137},  696  (1932).

\bibitem{kaza-etal:85}
A.~P. Kazantsev, G.~A. Ryabenko, G. Surdutovich, and V. Yakovlev, Phys. Rep.
  {\bf 129},  75  (1985).

\bibitem{johnson:77}
B.~R. Johnson, J. Chem. Phys. {\bf 67},  4086  (1977).

\end{thebibliography}

\newpage
\begin{table}
\begin{tabular}{@{}lll}
Parameter   & Rb & He* \\ \hline
$\lambda $ & $780\times 10^{-9} {\rm m}$ & $1083\times 10^{-9} {\rm m}$\\
$\Gamma$ & $2\pi 5.9\times 10^6 {\rm s}^{-1}$ & 
$2\pi 1.6\times 10^6 {\rm s}^{-1}$\\
$E_r$ & $6.4\times 10^{-4} \hbar \Gamma$ & $2.6\times 10^{-2} \hbar \Gamma$\\ 
$C^{(3)}$ & $0.113 \hbar \Gamma / \lambda ^3$ & 
$0.125 \hbar \Gamma / \lambda ^3$\\
state i & $5S_{1/2},F=2,m_F=2$ & $2^3S_1,J=1,m_J=1$\\
state f & $5S_{1/2},F=2,m_F=1$ & $2^3S_1,J=1,m_J=0$\\
$C_{ii}$ & $|E_{-1}(0)|^2 1/3 $ & $|E_{-1}(0)|^2 1/6$\\
$C_{ff}$ & $|E_{-1}(0)|^2 1/2 $ & $|E_{-1}(0)|^2 1/2 $\\
$C_{if}$ & $E_{-1}(0)^*E_0(0) \sqrt{2}/3 $ & $E_{-1}(0)^*E_0(0) 1/3 $\\
\end{tabular}
\caption{Values of parameters used in the text for Rb and He*: laser
wavelength $\lambda $, atomic transition width $\Gamma$, recoil
energy $E_r$, van der Waals coefficient
$C^{(3)}$ (see Eq.\protect\ref{eq:VLJ}).
Definition of the initial and final atomic state ($i$ and $f$).
Optical potential coefficients $C_{ij}$ (see
Eq.\protect\ref{eq:Vdip2}). We assume here that the $\sigma^-$ wave ($E_{-1}$) is
much stronger than the $\pi$ wave ($E_{0}$). The dielectric constant
of the wall is $\epsilon=3.49$ and $\kappa=1.52 k $.}
\label{tab:1}
\end{table}
%\newpage

%\clearpage
%\newpage

%\newpage

\newpage

%\newpage

%\newpage

\end{document}